\documentclass[twocolumn,american,english,aps,prl]{revtex4}
\usepackage[]{fontenc}
\usepackage[latin9]{inputenc}
\usepackage{array}
\usepackage{amsmath}
\usepackage{graphicx}
\usepackage{amssymb}

\makeatletter

\providecommand{\tabularnewline}{\\}

\@ifundefined{textcolor}{}
{%
 \definecolor{BLACK}{gray}{0}
 \definecolor{WHITE}{gray}{1}
 \definecolor{RED}{rgb}{1,0,0}
 \definecolor{GREEN}{rgb}{0,1,0}
 \definecolor{BLUE}{rgb}{0,0,1}
 \definecolor{CYAN}{cmyk}{1,0,0,0}
 \definecolor{MAGENTA}{cmyk}{0,1,0,0}
 \definecolor{YELLOW}{cmyk}{0,0,1,0}
 }

\usepackage{bm}

\makeatother

\usepackage{babel}

\makeatother

\usepackage{babel}

\makeatother

\usepackage{babel}

\makeatother

\usepackage{babel}

\makeatother

\usepackage{babel}

\makeatother

\usepackage{babel}

\makeatother

\usepackage{babel}

\begin{document}

\title{Magnetic impurities in the honeycomb Kitaev model}

\author{Kusum Dhochak$^{1}$, R. Shankar$^{2}$ and V. Tripathi$^{1}$}

\affiliation{$^{1}$Department of Theoretical Physics, Tata Institute of Fundamental
Research, Homi Bhabha Road, Navy Nagar, Mumbai - 400005, India}

\email{vtripathi@theory.tifr.res.in}

\affiliation{$^{2}$Institute of Mathematical Sciences, C. I. T. Campus, Taramani,
Chennai - 600113, India}
\begin{abstract}
We study the effect of coupling magnetic impurities to the honeycomb
lattice spin$-1/2$ Kitaev model in its spin liquid phase. We show
that a spin-$S$ impurity coupled to the Kitaev model is associated
with an unusual Kondo effect with an intermediate coupling unstable
fixed point $K_{c}\sim J/S$ separating topologically distinct sectors
of the Kitaev model. We also show that the massless spinons in the
spin liquid mediate an interaction of the form $S_{i\alpha}^{2}S_{j\beta}^{2}/R_{ij}^{3}$
between distant impurities unlike the usual dipolar RKKY interaction
$S_{i\alpha}S_{j\alpha}/R_{ij}^{3}$ noted in various 2D impurity
problems with a pseudogapped density of states of the spin bath. Furthermore,
this long-range interaction is possible only if the impurities (a)
couple to more than one neighboring spin on the host lattice and (b)
the impurity spin $S\neq1/2.$ 
\end{abstract}
\maketitle
Impurity effects are an essential part of our understanding of strongly
correlated electron systems , both as a probe for the underlying electronic
state as well as due to the numerous nontrivial effects they have
on the properties of the system \cite{gegenwart08,schroder00}. Recently
many studies have been made of impurity effects as a probe for the
putative quantum spin liquid state in underdoped cuprate superconductors
\cite{khaliullin97} and geometrically frustrated magnets \cite{kolezhuk06}.
The $S=1/2$ honeycomb lattice Kitaev model \cite{kitaev03} provides
a very appealing playground in this context - it has a gapless spin
liquid phase and short range spin correlations \cite{baskaran07}
making it different from many other extensively studied spin liquids
\cite{khaliullin97,kolezhuk06,hermele05}; and crucially, the model
is integrable via several schemes of spin-fractionalization into fermions
\cite{kitaev03,feng07}. The Kitaev model has been
studied in various contexts ranging from the possibility of quantum
computation with the anyons \cite{kitaev03,dusuel08} that the model
predicts, understanding dynamics of quantum quenches in a critical
region \cite{krishnendu08} to fractional charge excitations in topological
insulators \cite{lee07}. However no study of magnetic impurity effects
in the Kitaev spin liquid has yet been made, which constitutes the
subject of this paper.

We study the behavior of spin-$S$ impurities in the gapless spin
liquid regime of the Kitaev model on the honeycomb lattice. The impurity
coupling $K$ scales away from an unstable fixed point $K_{c}\sim J/S$
irrespective of the sign of impurity coupling, similar to impurity
problems in pseudogapped bosonic spin liquids \cite{florens06}. The
Kitaev magnetic impurity problem is nevertheless qualitatively different
for two important reasons. First, as we show below, the unstable fixed
point separates topologically distinct sectors in the Kitaev model,
with the strong coupling sector associated with non-abelian anyons.
Second, the gapless spinons in the Kitaev spin liquid mediate a non-dipolar
RKKY interaction proportional to $S_{i\alpha}^{2}S_{j\beta}^{2}/R_{ij}^{3}$
between distant magnetic impurities provided that (a) each impurity
couples to more than one lattice site on the host and (b) the impurity
spin $S\neq1/2.$ The absence of long-range interaction for $S=1/2$
impurities opens a way for local manipulation of the Kitaev system.
A comparison of Kondo effect and RKKY interaction in graphene \cite{withoff90,saremi07},
a bosonic spin bath \cite{florens06} and the Kitaev model are shown
in Table \ref{tab:comparison}.

\begin{table*}
\begin{tabular}{|>{\raggedright}p{0.9in}|>{\raggedright}p{1.9in}|>{\raggedright}p{2.1in}|>{\raggedright}p{1.9in}|}
\hline 
 & Graphene  & $Z_{2}$ bosonic spin bath with pseudogap density of states $\rho(\epsilon)=C|\epsilon|.$  & Kitaev, honeycomb lattice\tabularnewline
\hline
\hline 
Kondo scaling  & Unstable intermediate coupling fixed pt. only for AFM coupling. Only
AFM flows to strong coupling above unstable fixed pt.  & Flow direction is independent of the sign of magnetic impurity coupling.
Unstable intermediate coupling fixed pt. for both FM and AFM.  & Scaling same as $Z_{2}$ bosonic spin bath case. However a topological
transition is associated with the unstable fixed point. \tabularnewline
\hline 
RKKY  & $S_{i\alpha}S_{j\alpha}/R_{ij}^{3}$  & $S_{i\alpha}S_{j\alpha}/R_{ij}^{3}$  & $S_{i\alpha}^{2}S_{j\beta}^{2}/R_{ij}^{3}$\tabularnewline
\hline
\end{tabular}

\caption{\label{tab:comparison}Comparison of Kondo effect and RKKY interaction
in graphene, a $Z_{2}$ bosonic spin bath with a pseudogap density
of states and the Kitaev model on the honeycomb lattice.}

\end{table*}

\begin{figure}
\begin{centering}
\includegraphics[height=2.9cm]{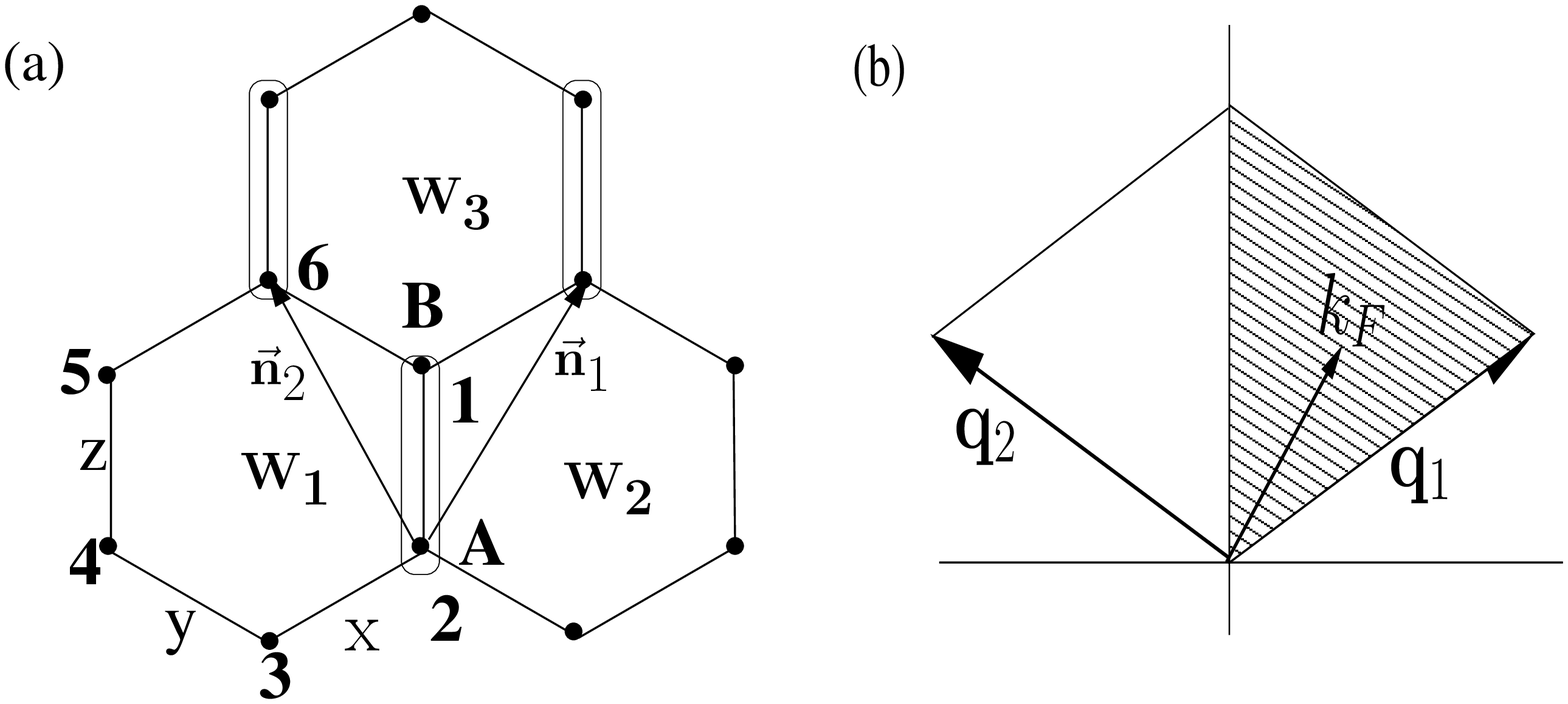} 
\par\end{centering}

\caption{\label{fig:basis}(a) Schematic of the Kitaev lattice showing the
$A$ and $B$ sites and the $x,$ $y$ and $z$ types of bonds. (b)
Figure showing the reciprocal lattice vectors for the $A$ sublattice.
The Dirac point for the massless Majorana fermions is denoted by $k_{F}$
and momentum summations are over the (shaded) half Brillouin zone.}

\end{figure}

The $S=1/2$ Kitaev model \cite{kitaev03} is a honeycomb lattice
of spins with direction-dependent nearest neighbor exchange interactions,
\begin{equation}
H_{0}=-J_{x}\!\!\!\sum_{\text{\ensuremath{x}-links}}\sigma_{j}^{x}\sigma_{k}^{x}-J_{y}\!\!\!\sum_{\text{\ensuremath{y}-links}}\sigma_{j}^{y}\sigma_{k}^{y}-J_{z}\!\!\!\sum_{\text{\ensuremath{z}-links}}\sigma_{j}^{z}\sigma_{k}^{z},\label{eq:Hamiltonian}\end{equation}
 where the three bonds at each site (see Fig.\ref{fig:basis}) are
labeled as $x,\, y$ and $z$. As was shown by Kitaev, the flux operators
$W_{p}=\sigma_{1}^{x}\sigma_{2}^{y}\sigma_{3}^{z}\sigma_{4}^{x}\sigma_{5}^{y}\sigma_{6}^{z}$
defined for each elementary plaquette $p$ are conserved (see Fig.\ref{fig:basis}),
with eigenvalues $\pm1,$ and form a set of commuting observables.
Each of the Kitaev spins is represented in terms of Majorana fermions
$b_{i}^{x},\, b_{i}^{y},\, b_{i}^{z},\, c_{i}$ as $\sigma_{i}^{\alpha}=ib_{i}^{\alpha}c_{i},$
which span a larger Fock space, and we restrict to the physical Hilbert
space of the spins by choosing the gauge \cite{kitaev03} $D_{i}=ib_{i}^{x}b_{i}^{y}b_{i}^{z}c_{i}=1.$
On each $\alpha-$type bond, $u_{ij}^{\alpha}=ib_{i}^{\alpha}b_{j}^{\alpha}$
is also conserved and the ground state manifold corresponds to a vortex
free state where all $W_{i}$ are equal. In the vortex free state,
we can fix all $u_{ij}=1$ (corresponds to $W_{p}=1$) and the Hamiltonian
can be written as a theory of non interacting Majorana fermions. The
reduced Hamiltonian for this ground state manifold is given by $H_{0}=\frac{i}{4}\sum_{jk}A_{jk}c_{j}c_{k},$
where $A_{jk}=2J_{{\alpha}_{_{jk}}}$ if $j,\, k$ are neighboring
sites on an $\alpha-$bond and zero otherwise. The excited state manifolds
(with finite vorticity) are separated from the ground state manifolds
by a gap of order $J_{\alpha}.$ Defining the Bravais lattice with
a two point basis (Fig.\ref{fig:basis}), the Hamiltonian can be diagonalized
in momentum space, $H_{0}=\frac{i}{4}\sum_{\mathbf{q}>0,\alpha}\epsilon_{\alpha}(\mathbf{q})a_{\mathbf{q},\alpha}^{\dagger}a_{\mathbf{q},\alpha},$
with $\epsilon_{\alpha}(\mathbf{q})=\pm|f(\mathbf{q})|,\:\: f(\mathbf{q})=2(J_{x}e^{ia\mathbf{q\cdot n_{1}}}+J_{y}e^{ia\mathbf{q\cdot n_{2}}}+J_{z})$
and the eigenbasis, $a_{\mathbf{q},0}=\tilde{c}_{\mathbf{q},A}+\tilde{c}_{\mathbf{q},B}e^{-i\tilde{\alpha}(\mathbf{q})}$
and $a_{\mathbf{q},1}=\tilde{c}_{\mathbf{q},A}-\tilde{c}_{\mathbf{q},B}e^{-i\tilde{\alpha}(\mathbf{q})}.$
Here $A/B$ is the site label for the two types of sites in Kitaev
model, $a$ is the lattice constant and $\tilde{\alpha}(\mathbf{q})$
is the phase of $f(\mathbf{q}).$ The sum over momenta is only over
half of first Brillouin zone. $\epsilon(\mathbf{q})$ has linear dispersion
around the Fermi point $\mathbf{k}_{F}$ (Fig.\ref{fig:basis}). For
simplicity, and without loss of generality, we henceforth assume $J_{x}=J_{y}=J_{z}=J.$

\textbf{Topological Kondo effect -} Consider a spin $S$ magnetic
impurity coupled to a Kitaev spin at an $A$ site ($\mathbf{r}=0$),
\begin{align}
V_{K} & =i\sum_{\alpha}K^{\alpha}S^{\alpha}b^{\alpha}c_{A}(0),\label{eq:Kondo-term}\end{align}
for which we perform a standard poor man's scaling analysis \cite{hewson92}
for the Kondo coupling $K.$ Consider the Lippmann-Schwinger expansion
for the $T-$matrix element, $\langle b^{\beta}|K^{\beta}S^{\beta}b^{\beta}c_{a,A}|(\mathbf{q},\alpha)\rangle$
(scattering of a $c-$Majorana with momentum $q$ and sublattice index
$\alpha$ to a $b-$Majorana), $T=T^{(1)}+T^{(2)}+\cdots,$ in increasing
powers of $K.$ %
\begin{figure}
\includegraphics[height=3.2cm]{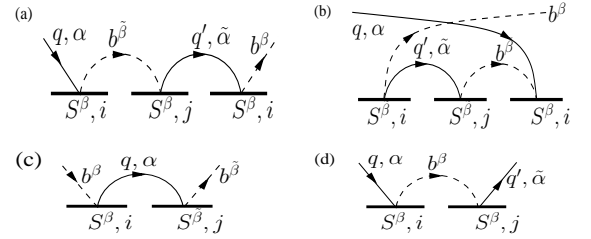} \caption{\label{fig:diag3}(a and b) Third order contributions to the $T-$matrix.
Site-diagonal scattering corresponds to $i=j$ and site off-diagonal
scattering, where relevant, corresponds to $i\neq j.$ Thin solid
lines correspond to $c-$Majoranas while dashed ones to $b-$Majoranas.
Thick solid lines represent the impurity spin. (c and d) New vertices
generated by off-diagonal scattering.}

\end{figure}
The first correction to the bare $T-$matrix comes from two \emph{third}
order terms (see Fig.\ref{fig:diag3}): $T^{(3)}\simeq-\frac{ia^{2}}{\sqrt{2}}K^{\beta}S^{\beta}\frac{\rho(D)\delta D}{JD}\sum_{\tilde{\beta}}(K^{\tilde{\beta}})^{2}(S^{\tilde{\beta}})^{2}.$
Here $\rho(\epsilon)=(1/2\pi v_{F}^{2})|\epsilon|\equiv C|\epsilon|$
is the density of states and $D$ is the band edge energy. If either
the impurity is a $S=\frac{1}{2}$ spin, or the Kondo interaction
is rotationally symmetric, the above contribution renormalizes the
Kondo coupling constant. However for $S\neq\frac{1}{2}$ with anisotropic
coupling, new terms are generated and one needs to go to higher order
diagrams to obtain the scaling of these new coupling terms.

Just as for the Kondo effect in graphene\cite{withoff90}, we also
need to consider the change in the density of states with bandwidth.
This gives a contribution $K\rightarrow K(D'/D)^{r},\quad(D'=D-|\delta D|).$
For $S=1/2$ or for symmetric impurity coupling we thus have \begin{align}
\delta K=-K\frac{\delta D}{D}\left(2K^{2}a^{2}CDS(S+1)/J-1\right).\label{eq:iso-kondo}\end{align}
The effective coupling $K$ has an unstable fixed point at $K_{c}=\sqrt{J/[2a^{2}\rho(D)S(S+1)]}\sim J/S.$
Here we used $D\lesssim J$ and $C\sim1/(Ja)^{2}.$ Clearly for $K>K_{c},$
the coupling flows to infinity independent of the nature of coupling
(ferromagnetic or antiferromagnetic), while for $K<K_{c},$ the coupling
flows to zero.

If the impurity couples to more than one Kitaev spin in a
plaquette, new contributions arise from site off-diagonal scattering ($i \ne j$ in Fig.\ref{fig:diag3} a, b). Adding all these contributions,
we find this also leads to a similar Kondo effect as was discussed
above for the single site coupling case. There are also new terms
of second order in $K$ that are generated (see Fig.\ref{fig:diag3}).
The term corresponding to Fig. \ref{fig:diag3}c is $\sim K^{\beta}K^{\tilde{\beta}}S^{\beta}S^{\tilde{\beta}}b_{i}^{\beta}b_{j}^{\tilde{\beta}}$.
When projected to the vortex free ground state, it becomes $\sim(K^{\beta})^{2}(S^{\beta})^{2}$
generating anisotropic potential for the impurity spin. The second
term, Fig.\ref{fig:diag3}d, is $\sim(K^{\beta})^{2}\:(S^{\beta})^{2}c_{i}c_{j}/J,$
which, as we shall see below, contributes to the long range interaction
among impurity spins.

A remarkable property of the Kondo effect in Kitaev model is that
the unstable fixed point is associated with a topological transition
from the zero flux state to a finite flux state. The strong coupling
limit amounts to studying the Kitaev model with a missing site or
cutting the three bonds linking this site to the neighbors. Kitaev
has shown \cite{kitaev03} that such states with an odd number of
cuts are associated with a finite flux, and also that these vortices
are associated with unpaired Majorana fermions and have non-abelian
statistics under exchange. Below we present another, perhaps more
physical, way of appreciating this result.

For the Hamiltonian $H=H_{0}+V_{K},$ the three plaquettes $W_{1},$
$W_{2}$ and $W_{3}$ that touch the impurity site are no longer associated with conserved 
flux operators, while the flux operators that do not include the origin remain conserved. The three plaquette operator $W_{0}=W_{1}W_{2}W_{3}$
is still conserved and $W_{0}=1$ in the ground state of the unperturbed
Kitaev model. The composite operators $\tau^{x}=W_{2}W_{3}S^{x},$
$\tau^{y}=W_{3}W_{1}S^{y}$ and $\tau^{z}=W_{1}W_{2}S^{z},$ where
the $S^{\alpha}$ are Pauli spin matrices corresponding to the impurity, are also conserved. The $\tau^{\alpha}$'s
do not mutually commute and instead obey the $SU(2)$ algebra, $[\tau^{\alpha},\tau^{\beta}]=2i\epsilon_{\alpha\beta\gamma}\tau^{\gamma}$
. This $SU(2)$ symmetry, which is exact for all couplings is realized
in the spin-1/2 representation $\left(\left(\tau^{\alpha}\right)^{2}=1\right)$.
Thus all eigenstates, including the ground state are doubly degenerate.

Consider a strong (antiferromagnetic) coupling limit $J_{K}\rightarrow\infty.$
The low energy states will be the ones in which the spin at the origin
forms a singlet $|0\rangle$ with the impurity spin, $|\psi\rangle=|\psi K^{-}\rangle\otimes|0\rangle.$ $|\psi K^{-}\rangle$ now stands for the low
energy states of the Kitaev model with the spin at the origin removed.
To see the action of the $SU(2)$ symmetry generators on these states,
we note that they can be written as $\tau^{\alpha}=\tilde{W}^{\alpha}\otimes\sigma^{\alpha}\otimes S^{\alpha}$
and $\tilde{W}^{\alpha}$ do not involve the components of the spin
at the origin, $\sigma^{\alpha}.$ We then have $\tau^{\alpha}|\psi\rangle=-(\tilde{W}^{\alpha}|\psi K^{-}\rangle)\otimes|0\rangle$.
Thus, in the strong coupling limit, the symmetry generators act nontrivially
only in the Kitaev model sector. This implies that the low energy
states of the Kitaev model with one spin removed are all doubly degenerate,
with the double degeneracy emerging from the Kitaev sector. This implies
there is a zero-energy mode in the single particle spectrum. The two
degenerate states correspond to the zero mode being occupied or unoccupied.
The same arguments for the double degeneracy in the Kitaev sector
may be repeated for the ferromagnetic strong coupling case.

Let us examine the structure of the zero mode. Removing a Kitaev spin
creates three unpaired $b-$Majoranas at the neighboring sites, say,
$b_{3}^{z},$ $b_{2}^{x}$ and $b_{1}^{y}.$ Note that $ib_{2}^{x}b_{1}^{y}$
is conserved and commutes with all the conserved flux operators $W_{i}$
but not with the two other combinations $ib_{1}^{y}b_{3}^{z}$ and
$ib_{3}^{z}b_{2}^{x}$ - then one can choose a gauge where the expectation
value of $ib_{2}^{x}b_{1}^{y}$ is equal to +1 so that these two $b$
modes drop out of the physics and we equivalently have one unpaired
$b-$Majorana. The unpaired $b_{3}^{z}$ Majorana has dimension $\sqrt{2};$
so there must therefore exist an unpaired Majorana mode in the $c$
sector (again of dimension $\sqrt{2}$) so that together these two
give the full (doubly degenerate) zero energy mode. We note that while
the $b_{3}^{z}$ mode is sharply localized, the wave function of the
$c$ mode can be spread out in the lattice.

That the ground state energy corresponding to a finite flux state
(pinned to the defect) is lower than the zero flux case in the Kitaev
model with a missing site has also been recently shown numerically\cite{willans10}.

\textbf{RKKY Interactions -} In the absence of impurities, in the
ground state manifold (vortex free state), we have only nearest neighbor
Kitaev spin correlations. This is because each Kitaev spin is a bilinear
of a massless $c$ Majorana and a massive $b$ Majorana, and the $b$
Majoranas have only short range correlations. Suppose we now add impurities
which may each be locally coupled to more than one Kitaev spin. Distant
impurities can interact only if they are coupled via the massless
Majoranas. By contracting $b$ Majoranas locally by second order perturbation
in the Kondo coupling, we generate vertices of the type $(K^{\alpha}S^{\alpha})^{2}c_{i}c_{j}/J,$
where $i$ and $j$ are not farther than nearest neighbor. Note that
since $c_{i}^{2}=1,$ these vertices will contain massless Majoranas
only when $i$ and $j$ belong to \emph{different} sites. This effectively
means that two distant impurities coupled to a single Kitaev site
each \emph{cannot} interact. However when the impurities interact
with more than one neighboring Kitaev spin, we shall see that a long
range interaction of the spins is possible. As an example, we analyze
the interaction when the two impurities are at the centers of distant
hexagons. The interaction term is $V_{K}=i\sum_{j\in\text{hex1},\alpha}K^{\alpha}S_{1}^{\alpha}b_{j}^{\alpha}c_{j}+i\sum_{j\in\text{hex2},\alpha}K^{\alpha}S_{2}^{\alpha}b_{j}^{\alpha}c_{j}.$%
\begin{figure}
\includegraphics[width=4.5cm]{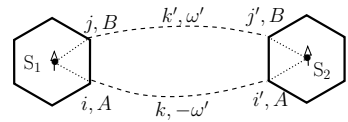} \caption{\label{fig:rkky1}Schematic of a typical long-distance impurity interaction
mediated by a pair of propagating $c-$Majoranas (shown by dashed
lines) emanating from the ends of a Kitaev bond. Another contribution
comes from crossed trajectories ($i,A\leftrightarrow j',B'$ and $j,B\leftrightarrow i',A'$).}

\end{figure}
The effective interaction generated involving $c-$type Majorana fermions
at two neighboring sites $(i\in A,\, j\in B)$ is given by (see Fig.\ref{fig:diag3}c,d)
$V_{\text{eff}}=\frac{2}{J}\sum_{a,<ij>_{\alpha}}(K^{\alpha_{ij}})^{2}(S_{a}^{\alpha_{ij}})^{2}c_{i}c_{j}.$
Here $\alpha_{ij}$ refers to the $z-$component when neighboring
sites $i$ and $j$ are along a $z-$bond, etc. Now the interaction
between the two impurity spins is given by the second order term in
$V_{\text{eff}}$ (or, equivalently, fourth order in $K$). These
terms are of the type $\frac{1}{J^{2}}\langle(K^{\alpha_{ij}})^{2}(S_{1}^{\alpha_{ij}})^{2}(K^{\beta_{i'j'}})^{2}(S_{2}^{\beta_{i'j'}})^{2}c_{i}c_{j}c_{i'}c_{j'}\rangle.$
Performing the fermionic averaging, the contribution (see Fig. \ref{fig:rkky1})
to the long range interaction from the pair of bonds $ij$ and $i'j'$
is \begin{align}
J_{12}^{ij,i'j'}\sim-(K^{\alpha_{ij}})^{2}(S_{1}^{\alpha_{ij}})^{2}(K^{\beta_{i'j'}})^{2}(S_{2}^{\beta_{i'j'}})^{2}\,\,\,\,\,\nonumber \\
\times\frac{a^{4}}{4v_{F}J^{2}\pi^{3}}\frac{1+\cos(2\tilde{\alpha}(\mathbf{k}_{F}))-2\cos(2\mathbf{k}_{F}\cdot\mathbf{R}_{12})}{R_{12}^{3}}.\label{eq:J12ij}\end{align}
Note that for spin-$1/2$ impurities, $(S^{\alpha})^{2}=1/4$ and
for isotropic coupling where $\sum_{\text{bond pairs}}(S_{1}^{\alpha_{ij}})^{2}(S_{2}^{\beta_{i'j'}})^{2}=\text{const.},$
no long-ranged interaction is generated. 

To summarize, we studied the effect of impurity quantum spins coupled
to the ground state manifold of the Kitaev model in the gapless spin-liquid
state. We found an unusual Kondo effect with an unstable fixed point
demarcating a topological transition between zero flux and finite
flux sectors. Where more than one impurity is present, we showed that
under certain circumstances, the massless spinons in the Kitaev model
mediate a higher order (non-dipolar) RKKY interaction between distant
impurity spins. The topological transition and the non-dipolar impurity
interaction make the Kitaev Kondo effect qualitatively different from
the Kondo effect in some bosonic spin liquids that also have an unstable
fixed point. We expect a similar scaling for the Kondo coupling and
RKKY interaction for other spin models \cite{yao09}with a similar
Majorana structure and phase diagram.

In the strong Kondo coupling limit we showed that a non-abelian anyon
is created consisting of an unpaired $b-$Majorana localized in its
vicinity and its delocalized $c-$counterpart. The localized $b-$Majorana
at the finite flux defect is very reminiscent of the localized Majoranas
in the cores of half vortices in $p-$wave superconductors \cite{ivanov01}.
One difference,
as pointed out by Kitaev, is that in $p-$wave superconductors, the
currents associated with the fluxes are charged (and thus interact
with impurity potentials). Besides, the individual charges acquire
abelian phases of their own. Another difference is that in the Kitaev
model, the full (doubly degenerate) zero mode is made of a $b$ and
a $c$ Majorana while in the superconductor, two vortex core Majoranas
make a zero mode %
\footnote{Vortex pairs can also be excited in the Kitaev model. See V. Lahtinen
\emph{et al.}, New J. Phys. \textbf{11}, 093027 (2009).%
}. $p-$wave paired ground states in the Kitaev model and its vortex
excitations have also been studied in Ref. \cite{yu08}.

We note that if we could adiabatically move the impurity spin (say
by making it on a STM tip), then we could move the anyons and thus
perform the braiding operations as is required for quantum computation.This method is more practical than perturbing an entire plaquette with terms like $\mu W_{i}$ that involve at least six Kitaev spins.

The temperature dependent impurity susceptibility can be measured
by NMR Knight shift experiments \cite{alloul75}. For weak Kondo coupling,
the magnetic susceptibility of the impurity is Curie-like with logarithmic
corrections, which flow to zero regardless of the sign of Kondo coupling.
In the strong coupling limit, we consider antiferromagnetic coupling for
$S=1/2$ impurities, where the anyons are vacancies in the Kitaev
lattice. The magnetic susceptibility has been shown \cite{willans10}
to have the form $\chi(T)\sim\frac{1}{T\ln(D/T)}$ for a pair of nearby
vacancies on the same sublattice while for a single vortex, $\chi(T)\sim\ln(D/T).$
In the absence of vacancies, the low temperature magnetic susceptibility
is small because of the spin gap in the Kitaev ground state. The nuclear
relaxation rate $T_{1}^{-1}$ may also be used as an impurity probe
\cite{dhochak09}. Using $\chi(T,\omega)\approx(\chi^{-1}(T,0)-i\omega)^{-1}$
and $T_{1}^{-1}\propto A^{2}I(I+1)T[\text{Im}\chi(T,\omega)/\omega]_{\omega\rightarrow0},$
where $A$ is the hyperfine coupling of spin$-I$ nuclei with the
defect, one gets $T_{1}^{-1}\sim T[\ln(D/T)]^{2}$ for the single
vortex and $T_{1}^{-1}\sim1/T[\ln(D/T)]^{2}$ for the two nearby defects
case, both of which qualitatively differ from the Korringa law $T_{1}^{-1}\sim T$
for the usual Kondo effect at low temperatures. There are already
numerous proposals in the literature how a Kitaev model could be realized
\cite{duan03}, and we are hopeful that eventually these novel impurity
effects may also be experimentally studied. 
\begin{acknowledgments}
The authors are grateful to G. Baskaran, K. Damle, D. Dhar and K.
Sengupta for valuable discussions during the course of this work.
K.D. and V.T. thank TIFR and R.S. thanks IMSc for support. V.T. also acknowledges DST for a Ramanujan
Grant {[}No. SR/S2/RJN-23/2006{]}.
\end{acknowledgments}

\end{document}